\pdfoutput=1
\documentclass[11pt,a4paper]{article}
\usepackage[margin=1in,headheight=14pt]{geometry}
\usepackage[T1]{fontenc}
\usepackage{lmodern,microtype,amsmath,amssymb,amsthm,mathtools}
\usepackage{booktabs,array,enumitem,xcolor,xurl}
\usepackage{fancyhdr}
\usepackage[unicode,colorlinks=true,linkcolor=blue!40!black,citecolor=blue!40!black,urlcolor=blue!45!black]{hyperref}
\hypersetup{pdftitle={Cohomological k-Equivalence Does Not Capture Graph Isomorphism},
  pdfauthor={Asterios Gkantzounis},
  pdfsubject={A negative answer to Question 14 of \'O Conghaile (MFCS 2022)},
  pdfkeywords={graph isomorphism, cohomological k-equivalence, Weisfeiler-Leman, integer relaxations, Cai-F\"urer-Immerman}}
\setlist{itemsep=2pt,topsep=3pt}

\newtheorem{theorem}{Theorem}
\newtheorem{lemma}[theorem]{Lemma}
\newtheorem{proposition}[theorem]{Proposition}
\theoremstyle{remark}\newtheorem*{remark}{Remark}

\newcommand{\F}{\mathbb{F}}
\newcommand{\Z}{\mathbb{Z}}

\DeclareMathOperator{\St}{St}
\DeclareMathOperator{\supp}{supp}
\DeclareMathOperator{\dom}{dom}
\DeclareMathOperator{\im}{im}
\DeclareMathOperator{\sgn}{sgn}
\DeclareMathOperator{\inv}{inv}
\newcommand{\eqZ}[1][k]{\equiv^{\Z}_{#1}}
\newcommand{\BF}{\textsf{BijForth}}
\newcommand{\val}{\mathrm{val}}

\title{Cohomological $k$-Equivalence Does Not Capture Graph Isomorphism\\[0.5em]
\large A negative answer to Question~14 of \'O~Conghaile}
\author{Asterios Gkantzounis\\[2pt]
\small School of Electrical and Computer Engineering\\
\small Technical University of Crete, Greece}
\date{17 November 2026}

\begin{document}
\maketitle
\thispagestyle{fancy}

\begin{abstract}
\'O~Conghaile introduced cohomological $k$-equivalence $\eqZ$, a refinement of equivalence in $k$-variable counting logic that is decidable in polynomial time for each fixed $k$: the local isomorphisms that survive must extend, in both directions, to compatible families of integer combinations of local isomorphisms. He asked whether, for every finite relational signature $\sigma$, some $k$ makes $\eqZ$ coincide with isomorphism of $\sigma$-structures (Question~14 of his MFCS~2022 paper). We show that the answer is negative already for graphs: for every $k\ge 1$ there are finite simple graphs $A_k\not\cong B_k$ with $A_k\eqZ B_k$.

Every isomorphism of the pair would induce a permutation $g$ of a common index set $D$. One part of the graphs, over $\F_3$, admits exactly the even permutations, because $g$ acts by its sign on a nonzero class in the cokernel of the unsigned boundary map on subsets of $D$. A second part, over $\F_2$ and built on the Johnson graph $J(|D|,2)$, admits exactly the odd permutations. No set of at most $k$ vertices detects the parity: every local isomorphism of a common family is realised, with the same images, by permutations of both parities. For each local isomorphism we build two compatible families of local distributions, one for each parity, whose denominators are powers of $2$ and of $3$ respectively. As in the Diophantine lower bound of Berkholz and Grohe, a B\'ezout combination yields integral extensions, while the nonnegative average yields the bijective forth property. The graphs have size doubly exponential in $k$.

We complement the main result with three observations. For the pair built from an index set of size $n$, the least width at which cohomological equivalence distinguishes the two graphs is $\log_4 n+O(1)$; so within this construction the doubly exponential size is unavoidable. The pairs are nevertheless distinguished in polynomial time by an explicit invariant that combines the class of the part over $\F_3$ with the parity of the part over $\F_2$. Finally, cohomological $2$-equivalence already distinguishes the coloured graphs with which Berkholz and Grohe show that no level of the Diophantine relaxation decides isomorphism.
\end{abstract}

\section{Introduction}

The Weisfeiler--Leman hierarchy, equivalently counting logic with a bounded number of variables, does not identify graphs up to isomorphism for any fixed number of variables, by the construction of Cai, F\"urer and Immerman~\cite{CFI92}. Their graphs encode systems of linear equations over $\F_2$, and later work, including~\cite{BG17,OC22,LP25}, has studied which algebraic strengthenings of local consistency overcome such examples. One of them is the Diophantine (integer) relaxation of isomorphism studied by Berkholz and Grohe~\cite{BG17}.

\'O~Conghaile~\cite{OC22,OClong}, see also Abramsky's notes~\cite{Abr22}, used the presheaf view of local consistency to define cohomological refinements: cohomological $k$-consistency for constraint satisfaction and cohomological $k$-equivalence $\eqZ$ for structure isomorphism. In the latter, the local isomorphisms that survive the $k$-pebble bijective game must also extend, in both directions, to compatible families of integer combinations of local isomorphisms (Definition~6 of~\cite{OC22}, recalled in Section~\ref{sec:cert}). For each fixed $k$ the relation is decidable in polynomial time~\cite[Section~4.3.2]{OC22}, see also~\cite[Section~16]{Abr22}, and it refines equivalence in $k$-variable counting logic~\cite[Proposition~16.1]{Abr22}. Moreover, there is a single $k$ such that, for every prime power $q$, $\eqZ$ coincides with isomorphism on all pairs $\mathrm{CFI}_q(G,g)$, $\mathrm{CFI}_q(G,h)$ with $\sum g=0$, where $\mathrm{CFI}_q$ is a Cai--F\"urer--Immerman-type construction over $\Z_q$ on an ordered base graph $G$~\cite[Theorem~12]{OC22}. Up to small differences in the encoding, this covers the CFI structures over the rings $\Z_{2^i}$ on totally ordered base graphs with which Lichter separated rank logic from polynomial time~\cite{Lic23}, as \'O~Conghaile points out~\cite{OC22}. In the same sense it covers the variant, with the same automorphisms, with which Dawar, Gr\"adel and Lichter showed that no fixed level of the invertible-map equivalences over all primes coincides with isomorphism~\cite{DGL23}. These examples therefore do not answer the following question. \'O~Conghaile asked~\cite[Question~14]{OC22} whether for every signature $\sigma$ there is a $k$ such that for all $\sigma$-structures $A,B$ we have
\[
A\cong B \iff A\eqZ B .
\]
The same question appears as Question~14 in the arXiv version of~\cite{OC22} and as Question~21 in the manuscript~\cite{OClong}. He added that, as the best known structure isomorphism algorithm is quasi-polynomial, he does not expect a positive answer, and that negative answers would help to understand the hard cases. We show that the answer is negative already for graphs.

\begin{theorem}[Main result]\label{thm:main}
For every integer $k\ge 1$ there are finite simple undirected graphs $A_k$, $B_k$ such that
\[
A_k\not\cong B_k \qquad\text{and}\qquad A_k\eqZ B_k .
\]
\end{theorem}

Here $\eqZ$ is the relation of Definition~6 of~\cite{OC22}, with the bijective forth property required for sections with fewer than $k$ elements, as in~\cite[Section~5.1]{OClong}; see the Remark in Section~\ref{sec:cert}. All pairs have the same signature, whose only symbol is the binary adjacency relation. The construction depends on $k$ while the signature stays fixed, which is what is needed to refute the quantifier ``there exists a $k$'' in Question~14.

\paragraph{Idea of the proof.}
Every isomorphism of the pair would induce a permutation $g$ of one index set $D$. The first part of the graphs, over $\F_3$, admits exactly the even $g$. The second, over $\F_2$, admits exactly the odd $g$. This rules out isomorphism. However, for every local isomorphism of at most $k$ vertices in the common family of Section~\ref{sec:family} we construct two compatible families of distributions, one for each sign. Their denominators are powers of two and of three, respectively. B\'ezout's identity then yields the integer coefficients required at the fixpoint of the cohomological test.

\paragraph{Further results.}
The graphs of Theorem~\ref{thm:main} have size doubly exponential in $k$. Theorem~\ref{thm:width} shows that this is intrinsic to the construction. For the pair $A(n)$, $B(n)$ built from an index set $D$ of size $n$, the least width at which $\equiv^\Z$ distinguishes the two graphs is $\log_4 n+O(1)$. Hence, within this family, the smallest pair that survives width $k$ has $2^{\Theta(4^k)}$ vertices. The pairs are not hard instances for isomorphism testing: Proposition~\ref{prop:inv} gives an invariant that is computable in polynomial time and takes different values on $A_k$ and $B_k$. As a consequence, a logic in which every sentence fails to distinguish $A_k$ from $B_k$ for some $k$ does not capture polynomial time. Finally, we return to the comparison with the Diophantine relaxation that \'O~Conghaile left open~\cite[Section~4.4]{OC22}. Already $\equiv^\Z_2$ distinguishes the coloured graphs with which Berkholz and Grohe show that no level of that relaxation decides isomorphism (Proposition~\ref{prop:bg}).

Section~\ref{sec:param} fixes the parameters. Section~\ref{sec:cert} reduces Definition~6 to a certificate (Lemma~\ref{lem:cert}). Section~\ref{sec:graphs} defines the graphs and proves that they are not isomorphic. Section~\ref{sec:ext} proves two extension lemmas, Section~\ref{sec:family} defines the common family of local isomorphisms, and Section~\ref{sec:dist} constructs its distributions. Section~\ref{sec:colours} transfers the same witnesses to uncoloured graphs. Section~\ref{sec:width} determines the width at which the construction is distinguished, Section~\ref{sec:invariant} gives the polynomial-time invariant, and Section~\ref{sec:bg} treats the graphs of Berkholz and Grohe. Section~\ref{sec:disc} discusses the relation to earlier work, the size of the graphs, and open questions. Appendix~\ref{app:checks} lists finite computational checks; they are not used in the proofs.

\paragraph{Relation to earlier work.}
Berkholz and Grohe proved that no level of the Diophantine relaxation decides isomorphism: for every $\ell$ there are non-isomorphic $3$-regular graphs of size $O(\ell)$ whose level-$\ell$ isomorphism system has an integral solution~\cite[Theorem~4.1]{BG17}. Two ingredients of the present proof come from their work. First, a linear system that has a $p$-solution and a $q$-solution for coprime $p,q$ (rational solutions whose nonzero values are powers of $p$, respectively of $q$) has an integral solution~\cite[Lemma~2.1]{BG17}; the same B\'ezout argument applies to rational solutions whose denominators are powers of $2$ and of $3$, which is the form used in Section~\ref{sec:dist}. Second, the characteristics $2$ and $3$ are combined: they use group constraint satisfaction problems over $\Z_2\times\Z_3$ and an or-construction. \'O~Conghaile observes that their test appears similar to checking the $\Z$-extendability of the empty local section, and leaves a full comparison as a direction for future work~\cite[Section~4.4]{OC22}. Compared with $\Z$-extendability of the empty section alone, Definition~6 demands more: every surviving local isomorphism, not only the empty one, must have integral extensions, and the bijective forth property must hold. Proposition~\ref{prop:bg} shows that this difference matters already at width two. On the side of constraint satisfaction, Lichter and Pago~\cite{LP25} combine systems of linear equations over $\Z_2$ and $\Z_3$ by an or-construction, together with the coprime-solution idea of~\cite{BG17}, to defeat several affine integer relaxations; for some of these relaxations the lower bound holds even when $k$ grows sublinearly with the size of the instance. They also express their first counterexample as graph isomorphism instances with bounded colour class size. They show that cohomological $k$-consistency solves the constraint satisfaction problem of their Theorem~1 for every $k\ge 4$, but fails for every constant $k$ on an NP-complete template obtained by modifying the or-construction~\cite[Theorem~3]{LP25}. Chan and Ng~\cite{CN25} proved the first lower bounds for cohomological $k$-consistency, for random instances of a class of constraint satisfaction problems. Conneryd, Ghannane and Pang~\cite{CGP26} derived level lower bounds for it from degree lower bounds for algebraic proof systems, for instance for $c$ versus $\ell$ colouring. These results concern constraint satisfaction. They do not by themselves give non-isomorphic structures that are $\eqZ$-equivalent. In the present construction the two characteristics are coupled through the parity of the permutation of $D$, and Lemma~\ref{lem:sign} shows that no local isomorphism of at most $k$ vertices decides this parity.

\section{Parameters and notation}\label{sec:param}

For the given $k\ge1$ we set
\[
d=\max\{k,3\},\qquad m=4^{d}+8(d+1)^{2}+1,\qquad n=2m,\qquad D=\{1,\dots,n\}.
\]
For a set $S$ and $j\notin S$ we write $S+j=S\cup\{j\}$. The choice of the parameters gives
\begin{align*}
&\text{(P1)}\quad n\ge 3d+4^{d}+2, &&\text{(P2)}\quad m>3\Bigl(d+\tbinom d2\Bigr)+3d+5,\\
&\text{(P3)}\quad n\ge 5d+4, &&\text{(P4)}\quad \min\{m,\,2(n-2)\}>4d+3 .
\end{align*}
(P1) leaves two interchangeable indices (Lemma~\ref{lem:sign}). (P2) allows modifications in the kernel of an incidence map (Lemma~\ref{lem:F3}). (P3) gives connectivity in Lemma~\ref{lem:F2}, and (P4) gives free variables in the local equations of Section~\ref{sec:dist}. In what follows we work at depth $k\le d$, so the same bounds also cover $k=1,2$. Only (P1)--(P4) are used. They already hold for $m=4^{d}+1$, so the term $8(d+1)^2$ is a safety margin only. For $k\le 3$ one may also take $m=40$, $n=80$: the right-hand sides of (P1) and (P2) are then $75$ and $32$, $n=80\ge19=5d+4$, and $\min\{40,156\}=40>15$. The number $n$ counts only the indices in $D$; the graphs contain, in addition, all value and row vertices introduced in Section~\ref{sec:graphs}.

\section{The certificate for Definition 6}\label{sec:cert}

Let $A$, $B$ be finite structures over the same relational signature and let $k\ge 1$. A \emph{partial isomorphism} from $A$ to $B$ is an isomorphism from an induced substructure of $A$ onto an induced substructure of $B$. For $C\subseteq A$ with $|C|\le k$, let $I_k(A,B)(C)$ be the set of partial isomorphisms with domain $C$. Restricting maps to subsets of $C$ turns $I_k(A,B)$ into a presheaf over the subsets of $A$ of size at most $k$. A \emph{subpresheaf} $\mathcal S$ is a family of such maps closed under restriction; $\mathcal S(C)$ denotes its elements with domain $C$. The inverse maps form the subpresheaf $\mathcal S^{-1}$ of $I_k(B,A)$.

We recall the definitions of~\cite[Sections~2.2, 4.2 and~4.3.2, Definition~6]{OC22}. The group $\Z\mathcal S(C)$ is the free abelian group with basis $\{[s]:s\in\mathcal S(C)\}$. For $C'\subseteq C$, restriction sends $[s]$ to $[s|_{C'}]$ and is extended linearly. For $t\in\mathcal S(C)$ the following properties are required.
\begin{enumerate}[label=(\arabic*)]
\item \emph{Integral extendability.} The section $t$ is $\Z$-extendable in $\mathcal S$ if there are $r_U\in\Z\mathcal S(U)$ for all $U\subseteq A$ with $|U|\le k$ such that $r_U|_{U'}=r_{U'}$ for every $U'\subseteq U$, and $r_C=[t]$. It is $\Z$-bi-extendable in $\mathcal S$ if, in addition, $t^{-1}$ is $\Z$-extendable in $\mathcal S^{-1}$.
\item \emph{Bijective forth.} If $|C|<k$, the property $\BF$ asks for a bijection $f:A\to B$ that extends $t$ and satisfies $t\cup\{a\mapsto f(a)\}\in\mathcal S$ for every $a\in A$.
\end{enumerate}
Let $F(\mathcal S)$ be the largest subpresheaf of $\mathcal S$ each of whose sections has the required properties with respect to $\mathcal S$. Starting from $I_k(A,B)$ we apply $F$ until it stabilises. Definition~6 of~\cite{OC22} writes $A\eqZ B$ when the fixpoint is nonempty, that is, when it contains the empty section.

\begin{remark}
In~\cite{OC22}, $\Z$-extendability asks for a $\Z$-linear global section $\{r_U\}$ over the cover of $A$ by its subsets of size at most $k$, that is, $r_U$ and $r_{U'}$ must agree after restriction to $U\cap U'$. Since the cover is closed under intersections, this is equivalent to compatibility under all inclusions, as in~(1). The bijective forth property is required only for sections with $|\dom t|<k$, as in the bijective $k$-pebble game and as stated in~\cite[Section~5.1]{OClong} and in~\cite[Section~15.3]{Abr22}. Read literally for all sections, it would fail for every section of size $k$ whenever $k<|A|$, because $I_k(A,B)$ contains no maps of size $k+1$; the fixpoint would then be empty for every pair with $|A|>k$, even for $A$ versus itself. For the family of Lemma~\ref{lem:cert}, the bijective forth property also provides sections over every set of at most $k$ vertices, so all readings of ``nonempty'' agree. Finally, the proof of Lemma~\ref{lem:cert} shows that its family satisfies (1) and (2) with respect to every subpresheaf of $I_k(A,B)$ that contains it. It therefore survives every iteration that only removes sections violating (1) or (2), such as the algorithm in~\cite[Section~5.1]{OClong}; compare also~\cite[Observation~10]{OClong}.
\end{remark}

\begin{lemma}\label{lem:cert}
Let $M$ be a nonempty family of partial isomorphisms of size at most $k$, closed under restriction. Suppose that for every $t\in M$ there are functions $Q=Q^t$ and $Z=Z^t$ on the set of all injective partial maps from $A$ to $B$ of size at most $k$, with nonnegative rational and with integer values respectively, such that
\[
Q_\emptyset=Z_\emptyset=Q_t=Z_t=1,\qquad \supp Z\subseteq \supp Q\subseteq M .
\]
For each $w\in\{Q,Z\}$, each partial isomorphism $u$ of size smaller than $k$, and all $a\in A\setminus\dom u$, $b\in B\setminus\im u$, we require
\begin{equation}\label{eq:marg}
\sum_{y\in B\setminus \im u} w_{u\cup\{a\mapsto y\}} \;=\; w_u \;=\; \sum_{a'\in A\setminus\dom u} w_{u\cup\{a'\mapsto b\}} .
\end{equation}
Each equality is imposed for every new element of the respective side, even when $w_u=0$. An extension that is not a partial isomorphism has coefficient zero (this follows from $\supp w\subseteq M$). Under these assumptions $A\eqZ B$.
\end{lemma}

\begin{proof}
Put $\mathcal S(C)=\{u\in M:\dom u=C\}$. Closure under restriction makes $\mathcal S\subseteq I_k(A,B)$ a subpresheaf. Fix $t\in M$ and write $C=\dom t$, $E=\im t$. On the source and on the target side, respectively, we define
\[
r_U=\sum_{u\in \mathcal S(U)} Z_u\,[u],\qquad r'_V=\sum_{\substack{u\in M\\ \im u=V}} Z_u\,[u^{-1}]\qquad(|U|,|V|\le k).
\]
If we remove one element $a$ from $U$, the restriction of $r_U$ lies in $\Z\mathcal S(U\setminus\{a\})$, because $M$ is closed under restriction. The coefficient of a basis element $[u]$ in it is the left-hand sum of~\eqref{eq:marg}, since the terms outside $M$ have coefficient zero by $\supp Z\subseteq M$; hence it equals $Z_u$. The right-hand equality gives the same for the restriction of $r'_V$ when one element of $V$ is removed. Successive removals prove compatibility for every inclusion.

Applying the same deletions to $Q$ down to the empty set, we obtain
\[
\sum_{\dom u=C} Q_u=Q_\emptyset=1,\qquad \sum_{\im u=E} Q_u=Q_\emptyset=1 .
\]
The term $Q_t=1$ exhausts both sums. By nonnegativity every other term is zero. The inclusion of supports and $Z_t=1$ therefore give $r_C=[t]$ and $r'_E=[t^{-1}]$. This proves that $t$ is $\Z$-bi-extendable in $\mathcal S$.

If $|C|<k$, we form the matrix $\Pi_{ab}=Q_{t\cup\{a\mapsto b\}}$ for $a\in A\setminus C$ and $b\in B\setminus E$. Every row and every column has sum $Q_t=1$; summing all entries shows that the two sides have the same number of elements. For every set $\Xi$ of rows, the mass $|\Xi|$ is distributed over the columns that have a positive entry in these rows, with capacity one per column. Hence there are at least $|\Xi|$ such columns. Hall's theorem~\cite{Hal35} gives a perfect matching inside the positive entries. Its union with $t$ is the required bijection $f$, because every chosen extension lies in $\supp Q\subseteq M$.

Finally, if $\mathcal S\subseteq\mathcal T\subseteq I_k(A,B)$, the same bijections and the same integral sections witness the properties of the elements of $\mathcal S$ also inside $\mathcal T$; for the integral sections it suffices to give the coefficient zero to the new basis elements. Hence $\mathcal S\subseteq F(\mathcal T)$. By induction from $\mathcal T=I_k(A,B)$, $\mathcal S$ is contained in every iterate and in the final fixpoint. $\mathcal S$ contains the empty section, so Definition~6 gives $A\eqZ B$.
\end{proof}

\section{The pair of graphs}\label{sec:graphs}

Let $D=[n]=\{1,\dots,n\}$ with $n=2m$. The graphs $A$ and $B$ are finite simple graphs with the same ten vertex colours
\[
D,\; D^*,\; P,\; O,\; H,\; X,\; L^0,\; L^1,\; L^2,\; Y .
\]
Each colour is interpreted by a unary predicate, and adjacency is one symmetric binary relation. (The colours are removed in Section~\ref{sec:colours}.) The indices and values below identify vertices inside a colour; the order of $[n]$ is used only to define edges. The two parts described next share the same set $D$, and their edges form the whole graph.

\paragraph{The part over $\F_3$.}
Let $V_j$ be the vector space over $\F_3$ with basis the $j$-subsets of $D$, with the standard bilinear form $\langle\cdot,\cdot\rangle$. The symmetric group on $D$ acts on $V_j$ by permuting the basis sets, without signs: $g\cdot S=gS$, so that $(gv)_{gS}=v_S$. For a basis element $T$ put $\partial T=\sum_{i\in T}(T\setminus\{i\})$; thus $(\partial h)_S=\sum_{T\in\St(S)} h_T$ for $h\in V_{m+1}$ and $|S|=m$, where $\St(S)=\{S+j:j\notin S\}$. Writing sets as squarefree monomials in variables $z_1,\dots,z_n$, we define
\[
\beta=\prod_{i=1}^{m}(z_{2i-1}-z_{2i})\in V_m .
\]
For $|T|=m+1$ and $a\in\F_3$ there is a vertex $L^\nu_{T,a}$ in each of the colours $L^\nu$, $\nu=0,1,2$, adjacent to all elements of $T$. We add the edges $L^0_{T,a}L^1_{T,a}$, $L^1_{T,a}L^2_{T,a}$ and $L^0_{T,a}L^2_{T,a+1}$. For $|S|=m$ and every vector $x\in\F_3^{D\setminus S}$ with $\sum_{j\notin S}x_j=\beta_S$ there is a vertex $Y_{S,x}$. Its neighbours are the elements of $S$ and the vertices $L^0_{S+j,x_j}$ for all $j\notin S$. This part is constructed in the same way in $A$ and in $B$.

Let $g$ be the permutation that an isomorphism induces on $D$. The neighbours in $D$ force the indices $gT$ and $gS$. Over each $T$, the vertex $L^0_{T,a}$ has exactly one neighbour in $L^1$, namely $L^1_{T,a}$, and one in $L^2$, namely $L^2_{T,a+1}$; so the nine vertices over $T$ form a $9$-cycle whose colour-preserving automorphisms are exactly its three rotations. In other words, the first two matchings between the colours $L^\nu$ force the same permutation of values on all three copies over $T$, and the third forces it to commute with the map $a\mapsto a+1$. Hence the image of the value $a$ is $a+h_{gT}$ for some $h_{gT}\in\F_3$. The image of $Y_{S,x}$ then necessarily has the value $x_j+h_{g(S+j)}$ at position $gj$. Summing these values we obtain
\begin{equation}\label{eq:F3}
\partial h=\beta-g\beta,\qquad\text{where } (g\beta)_{gS}=\beta_S .
\end{equation}
Conversely, every solution defines through these formulas bijections of the value and row fibres that preserve exactly the listed edges.

The class $[\beta]$ in $V_m/\partial V_{m+1}$ is nonzero. Indeed, the coefficient of $\partial^{\mathsf T}\beta$ at every $(m+1)$-set is zero: for a deletion to contribute, the set must contain both elements of one pair and one element of every other pair, and then exactly two deletions contribute, with opposite signs. Hence $\beta$ is orthogonal to the image of $\partial$, while $\langle\beta,\beta\rangle=2^m\neq0$ in $\F_3$.

Every transposition inside an original pair $\{2i-1,2i\}$ sends $\beta$ to $-\beta$. For a transposition between two pairs we write the corresponding factors, up to a common sign, as $(a-b)(c-d)$ so that $b$ and $c$ are exchanged. In $\F_3$ the sum of this product and its image is
\[
ab+ac+ad+bc+bd+cd=\tfrac12\,\partial(abc+abd+acd+bcd).
\]
On monomials with disjoint sets of variables $\partial$ satisfies the product rule. Since the boundary of each of the remaining factors $z_{2l-1}-z_{2l}$ is zero, the same sum for the whole of $\beta$ lies in the image of $\partial$. Transpositions generate the symmetric group, and $\partial$ commutes with its action. Therefore $g[\beta]=\sgn(g)[\beta]$. Since $2[\beta]\neq0$, equation~\eqref{eq:F3} has a solution exactly when $g$ is even.

\paragraph{The part over $\F_2$.}
Let $J=J(n,2)$ be the Johnson graph on the $2$-subsets of $D$, with $e\sim f$ when $|e\cap f|=1$, and write $N(e)$ for the neighbours of $e$ in $J$. We join every $i\in D$ to its copy $i^*\in D^*$. For every $e$ we add a vertex $P_e$ adjacent to the two elements of $e$. For every $i\neq j$ the vertex $O_{ij}$ is adjacent to $i$, $j^*$ and $P_{\{i,j\}}$. For every unordered edge $\{e,f\}$ of $J$ and every $b\in\F_2$ the vertex $H_{ef,b}=H_{fe,b}$ is adjacent to $P_e$ and $P_f$.

For every $e=\{i,j\}$ with $i<j$ and every $x\in\F_2^{N(e)}$ we add a vertex $X_{e,x}$, adjacent to $P_e$ and to $H_{ef,x_f}$ for every $f\sim e$. The vertex $X_{e,x}$ has one more neighbour: $O_{ij}$ (increasing order) if $\sum_f x_f=c_e$, and $O_{ji}$ (decreasing order) otherwise. In $A$ we have $c_e=0$ for all $e$. In $B$ we use values $c'_e$ whose only nonzero value is $c'_{\{1,2\}}=1$.

The permutation $g$ determines the images in $D^*$, $P$ and $O$. Over every edge $ef$ the map on the values of $H$ has the form $b\mapsto b+a_{ef}$ with $a_{ef}=a_{fe}$. The edges to $H$ force the image of $X_{e,x}$ to have the coordinate $x_f+a_{ef}$ at position $gf$. For $e=\{i,j\}$ with $i<j$ put $\epsilon_e(g)=1$ if $g(i)>g(j)$ and $\epsilon_e(g)=0$ otherwise. Preserving the ordered neighbour is equivalent to
\begin{equation}\label{eq:F2}
\sum_{f\sim e}a_{ef}=c_e+c'_{ge}+\epsilon_e(g)\qquad(e\in V(J)).
\end{equation}
Every solution gives a full isomorphism of this part, since the permutation of positions and the translations of values are bijective on every fibre. Summing over $e$, every variable appears twice, while the right-hand side gives $1$ plus the number $\inv(g)$ of inversions of $g$, modulo $2$. Conversely, every demand vector with sum zero is realised: we take a spanning tree of the connected graph $J$, set the values outside the tree to zero, and satisfy the equation of each leaf in turn through the edge to its parent. At the root the last requirement is satisfied because the total sum is zero. Hence~\eqref{eq:F2} has a solution exactly when $g$ is odd. The same $g$ must act on the common set $D$ of both parts, and the opposite parity requirements prove $A\not\cong B$.

\section{Extending prescribed shifts}\label{sec:ext}

For an $m$-set $S$ the \emph{star} is $\St(S)=\{S+j:j\notin S\}$; its elements are called columns and $S$ is its centre. The image of a row vertex $Y_{S,x}$ prescribes the shifts on the whole star $\St(gS)$, while the image of a single value vertex prescribes only one shift. So, even when the global equation is solvable, we need a solution with the given local values.

\begin{lemma}[Extension over $\F_3$]\label{lem:F3}
Recall that $n=2m$. Suppose that $\partial h=b$ has a solution $h\in V_{m+1}$. Prescribe the values of $h$ on all columns of at most $r$ stars and on at most $r$ further columns. If the prescriptions agree where they meet and have sum $b_S$ on every prescribed star $\St(S)$, then some solution realises them all, provided
\[
m>3\Bigl(r+\tbinom r2\Bigr)+3r+5 .
\]
\end{lemma}

\begin{proof}
We subtract one solution. It now suffices to construct an element of $\ker\partial$ with the required values, which have sum zero on every marked star. For every $(m+3)$-set $\Gamma$, $\partial^2\Gamma$ has coefficient $2$ at each $(m+1)$-subset of $\Gamma$ and zero elsewhere. Moreover $\partial^3\Gamma=0$, because each set of three deletions is counted $3!$ times. Hence $\tfrac12\partial^2\Gamma\in\ker\partial$, and it has coefficient $1$ at every $(m+1)$-subset of $\Gamma$.

Let $W$ be the set of the separately prescribed columns together with all columns that belong to two different marked stars. Two marked stars have at most one common column, the union of their centres, so $|W|\le r+\binom r2$.

\emph{Step 1.} For each $T\in W$ we choose distinct $\alpha,\alpha'\in D\setminus T$ such that $\alpha,\alpha'\notin T'\setminus T$ for every $T'\in W\setminus\{T\}$ with $|T'\setminus T|\le2$. At most $2(|W|-1)$ elements are excluded, and $n-m-1=m-1$ elements lie outside $T$; since $m-1-2(|W|-1)\ge2$ by the hypothesis, the choice is possible. No other column $T'\in W$ is contained in $T\cup\{\alpha,\alpha'\}$, because such a column would satisfy $1\le|T'\setminus T|\le2$ and $T'\setminus T\subseteq\{\alpha,\alpha'\}$. A suitable multiple of $\tfrac12\partial^2(T\cup\{\alpha,\alpha'\})$ therefore corrects the value at $T$ without changing the other values on $W$.

\emph{Step 2.} Once all values on $W$ are adjusted, the remaining difference on each marked star $\St(S)$ has sum zero, since the current element lies in $\ker\partial$; it is supported on the private columns of the star, that is, those not in $W$. For two private columns $S+j$ and $S+l$ we choose distinct $\alpha,\alpha'\notin S\cup\{j,l\}$ such that $\alpha,\alpha'\notin R\setminus S$ for every other marked centre $R$ with $|R\setminus S|\le3$, and $\alpha,\alpha'\notin T'\setminus S$ for every $T'\in W$ with $|T'\setminus S|\le3$. At most $2+3(r-1)+3(r+\binom r2)$ of the $n-m=m$ elements outside $S$ are excluded, and the hypothesis leaves at least two of them available. The element
\[
\tfrac12\,\partial^2\bigl(S\cup\{j,\alpha,\alpha'\}\bigr)-\tfrac12\,\partial^2\bigl(S\cup\{l,\alpha,\alpha'\}\bigr)
\]
lies in $\ker\partial$ and gives on $\St(S)$ the difference of the two corresponding basis vectors: the terms at the columns $S+\alpha$ and $S+\alpha'$ cancel. If the first upper set contained another marked centre $R$, we would have $R\setminus S\subseteq\{j,\alpha,\alpha'\}$ with $|R\setminus S|\le3$, hence $R\setminus S=\{j\}$ by the choice of $\alpha,\alpha'$; then $S+j=R\cup S$ would be a common column of two marked stars, which is not private. The same holds for $l$. Nor can a column $T'\in W$ be contained in one of the two upper sets: then $T'\setminus S$ would be contained in $\{j,\alpha,\alpha'\}$ or in $\{l,\alpha,\alpha'\}$, hence equal to $\{j\}$ or $\{l\}$, so $T'$ would coincide with one of the two private columns. Consequently the corrections preserve every other prescribed star and the values on $W$. The pairwise differences generate all vectors of sum zero on the private columns. This completes the construction.
\end{proof}

\begin{lemma}[Extension over $\F_2$]\label{lem:F2}
In $J(n,2)$ let $(b_e)$ be divergence demands with $\sum_e b_e=0$, that is, we seek $a\in\F_2^{E(J)}$ with $\sum_{f\sim e}a_{ef}=b_e$ for all $e$. Prescribe the values $a_{ef}$ on all edges incident with $s$ chosen centres and on $q$ further edges. If the prescriptions agree on common edges and give the correct divergence at each of the $s$ centres, they extend to a global solution whenever $n\ge 2s+3q+4$.
\end{lemma}

\begin{proof}
We remove the $s$ centres and the remaining prescribed edges. At the remaining vertices we subtract the known incident values from the demand. The new total demand is zero, by the equations at the removed centres and because every prescribed edge is counted twice.

The removed objects involve at most $2s+3q$ indices of $D$. We choose four other indices. The Johnson subgraph on their pairs is connected and untouched. From every surviving pair we reach it in at most two steps: each member that is not among the four indices is replaced by one of the four indices that differs from both current members. Every new vertex and edge on the way contains one of the four indices and therefore survived. On the connected remainder we set zero outside a spanning tree and solve the equations at the leaves in turn. The last equation is satisfied because the total sum is zero.
\end{proof}

\section{A common family of partial isomorphisms}\label{sec:family}

For every colour, a \emph{fibre} is the set of vertices with the same index: for example all $Y_{S,x}$ with the same $S$, or all $H_{ef,b}$ with the same edge $ef$. In $D$, $D^*$, $P$, $O$ every fibre is a singleton. A permutation $g$ of $D$ determines bijectively how fibres are transported. Moving a vertex inside its fibre also needs the values listed in Table~\ref{tab:shifts}.

\begin{table}[t]
\centering
\begin{tabular}{lll}
\toprule
Source vertex & Shifts it reads & Target fibre\\
\midrule
$D,\ D^*,\ P,\ O$ & none & determined by $g$\\
$L^\nu_{T,b}$ & $h_{gT}$ & $(L^\nu,\,gT)$\\
$Y_{S,x}$ & all $h_{gT}$ for $T\in\St(S)$ & $(Y,\,gS)$\\
$H_{ef,b}$ & $a_{ef}$ & $(H,\,\{ge,gf\})$\\
$X_{e,x}$ & all $a_{ef}$ for $f\sim e$ & $(X,\,ge)$\\
\bottomrule
\end{tabular}
\caption{Shifts read by each source vertex, and its target fibre.}\label{tab:shifts}
\end{table}

Let $u$ be a partial map of at most $k$ vertices. We put $u$ in $M$ if there are a permutation $g$ and values of all shifts read by the domain of $u$ (Table~\ref{tab:shifts}), common wherever the variables coincide, with the following properties.
\begin{enumerate}[label=\arabic*.]
\item The image of every vertex is the one defined by $g$ and the transfer formulas of the two parts. In particular the values at the coordinates $gj$ and $gf$ of the images of the two kinds of rows are $x_j+h_{g(S+j)}$ and $x_f+a_{ef}$, respectively.
\item At every observed centre $S$ of a row $Y$, equation~\eqref{eq:F3} holds at position $gS$. At every observed centre $e$ of a row $X$, equation~\eqref{eq:F2} holds at position $e$. At centres where no row is read, no equation is imposed.
\end{enumerate}
Here the shifts of the ternary part are indexed by target columns, $h_{gT}$, and those of the binary part by source edges, $a_{ef}$. We call such $g$ and values a \emph{local realisation} of $u$. The empty map belongs to $M$, and every restriction inherits the data of the same realisation.

Every $u\in M$ is a partial isomorphism. Colours are preserved by construction, and $g$ preserves all incidences between indices; this settles every pair of vertices whose adjacency depends only on colours and indices. The remaining cases are the following.
\begin{itemize}
\item Inside a fibre, the rearrangement and the common translation of coordinates preserve the distinctness of vertices, so $u$ is injective.
\item $X_{e,x}$ and $H_{ef,b}$ are adjacent exactly when $b=x_f$, and both images are shifted by the common value $a_{ef}$.
\item $Y_{S,x}$ and $L^0_{S+j,a}$ are adjacent exactly when $a=x_j$, and both images are shifted by the common value $h_{g(S+j)}$.
\item Over the same $T$, the images of $L^0$, $L^1$, $L^2$ are shifted by the common value $h_{gT}$, which preserves and reflects the two identity matchings and the step $a\mapsto a+1$.
\item For an observed $X_{e,x}$, equation~\eqref{eq:F2} at $e$ preserves the choice between the neighbours $O_{ij}$ and $O_{ji}$.
\item The image of an observed row $Y_{S,x}$ is a vertex: by~\eqref{eq:F3} at $gS$ it satisfies the sum condition of the target.
\end{itemize}
This controls the edges and the non-edges between any two vertices of the domain.

\begin{lemma}[Sign change with the same prescribed images]\label{lem:sign}
Let $n\ge 3k+4^k+2$. For every $t\in M$, $|t|\le k$, there are local realisations of the same $t$ by an even and by an odd permutation of $D$.
\end{lemma}

\begin{proof}
We start from a realising permutation $g$. For a vertex of $D$ or of its copy $D^*$ we protect its index. For a pair or an ordered pair ($P$ or $O$) we protect both ends. For a value on an edge of the Johnson graph ($H$) we protect its three indices, and for a binary row at $e=\{i,j\}$ the two ends $i,j$. In total at most $3k$ indices are protected.

Every vertex of the domain partitions the other indices into at most four classes. For a binary row with values $x$ we record, for each $l\notin e$, the pair $(x_{\{i,l\}},x_{\{j,l\}})$. For a row $(S,x)$ over $\F_3$ we record whether $l\in S$ and, otherwise, the value $x_l$. For a single ternary value or one of its copies, membership in $T$ suffices. For the remaining vertices the protected indices suffice. Hence the at least $n-3k\ge 4^k+2$ unprotected indices have at most $4^k$ different joint profiles, and two of them have the same profile. Let $\tau$ be their transposition.

Since $\tau$ fixes every protected index and exchanges two indices with the same profile, it fixes every observed index set $T$ or $S$ setwise and every observed binary centre pointwise, and every observed row satisfies $x\circ\tau=x$: for a row over $\F_3$ the two exchanged values coincide, and for a binary row $(x_{\{i,l\}},x_{\{j,l\}})$ is the same for the two exchanged indices $l$. In particular $\tau$ does not reverse the order of the ends of any observed binary centre, so $\epsilon_e(g\tau)=\epsilon_e(g)$ and $c'_{g\tau e}=c'_{ge}$. In the ternary part we keep the same shifts at the same target columns when we replace $g$ by $g\tau$: since $\tau$ fixes the observed sets, every vertex reads the same target columns, the right-hand side of~\eqref{eq:F3} at $g\tau S=gS$ is unchanged, and the values of a source row agree at the positions that exchanged names. In the binary part, where the shifts are written in source coordinates, we relabel them by the global relabelling of source edges $a'_{ef}=a_{\tau e,\tau f}$; on an observed star this reads $a'_{ef}=a_{e,\tau f}$. It is automatically consistent on common edges, and it keeps the single values that were defined, since their indices are protected. The local equations at the observed rows still hold, and all prescribed images of $t$ remain exactly the same. The permutations $g$ and $g\tau$ have opposite parity.
\end{proof}

\section{Two distributions for each local map}\label{sec:dist}

We fix an arbitrary $t\in M$, including $t=\emptyset$. Lemma~\ref{lem:sign} gives an even and an odd realisation of the same $t$. We keep separately the permutation and the values prescribed by each realisation.
\begin{itemize}
\item For the even $g$, equation~\eqref{eq:F3} has a global solution. By Lemma~\ref{lem:F3} we choose a solution that agrees with $t$ on all its shifts. We use the corresponding global map on the ternary part and local distributions over $\F_2$ on the binary part.
\item For the odd $g$, equation~\eqref{eq:F2} has a global solution. By Lemma~\ref{lem:F2} we choose a solution that agrees with $t$. We use this global map on the binary part and local distributions over $\F_3$ on the ternary part.
\end{itemize}
In every application of the lemmas there are at most $|t|\le k$ prescribed centres and at most $|t|$ additional variables, so (P2) and (P3) apply with $r,s,q\le k\le d$. In both branches $g$ determines the action on the singleton fibres.

\paragraph{Compatibility of the local distributions.}
We consider one branch and write $p$ for the characteristic of the part on which we use local distributions. For every set $U$ of source vertices with $|U|\le k$ we read the variables of the whole $U\cup\dom t$ as in Table~\ref{tab:shifts}. On their union we choose uniformly a solution of the following system over $\F_p$. Every shift read by $t$ has its prescribed value. At every centre whose row is observed, the corresponding sum equation holds. In the ternary part the columns are written in target indices as in~\eqref{eq:F3}.

Every ternary star has $m$ variables and every binary star has $2(n-2)$. Two different stars in the same part meet in at most one variable. Put $v=k+|t|\le 2d$. At most $v$ stars and at most $v$ single variables occur in the union. Hence every star whose centre was not prescribed by $t$ has at least one variable that is private with respect to all other stars and single variables; for this $\min\{m,2(n-2)\}\ge 2v$ suffices, and (P4) gives $\min\{m,2(n-2)\}>4d+3\ge 2v+3$. We choose one such variable in every star that is not prescribed; they are all distinct. We set the other free variables arbitrarily and solve each equation for its private variable. The equations at the centres of $t$ already hold. Hence the space of choices is nonempty.

When we pass from $U$ to $U\cup\{a\}$, every admissible assignment of the old space extends. A new single variable is either already known or free: an observed row reads its whole star, so a single variable outside the old space lies in no star of the old system and occurs in no old equation. A row at a known centre reads the same variables. At a new centre we choose a private variable outside the old space and solve the one new equation. The restriction maps between the affine solution spaces are therefore onto, and all fibres of each of them have the same size, the order of the kernel of the corresponding linear map. The uniform distribution restricts exactly to the uniform distribution of the old space. The probabilities of all events have denominators that are powers of $p$.

\paragraph{Equations on the source and on the target side.}
For every partial map $u$ we define $R_p(u)$ as the probability, under the distribution for $U=\dom u$, that all vertices of $U$ receive the images given by $u$. We use the global map on one part, the shifts of the sample on the other, and the same fixed $g$ everywhere. Every outcome belongs to $M$. Moreover $R_p(\emptyset)=R_p(t)=1$.

For a new source element $a$, in every sample for $U\cup\{a\}$ the vertex $a$ has exactly one image. Under the event $u$ this image lies outside $\im u$, by the injective action on each fibre and the distinctness of the fibres. The compatibility of the distributions therefore proves the left-hand equality of~\eqref{eq:marg}.

For a new target element $b$, the fixed $g$ determines the one source fibre that can contain a preimage of $b$. For every candidate preimage $a$ in this fibre, the variables, the observed centres and the equations for $U\cup\{a\}$ are the same; they depend on the fibre and not on the vertex $a$. We therefore use the same sample for all candidates. Every such sample gives a bijection of the whole fibre. For the rows $Y$ this holds because the sum equation sends the affine hyperplane of source values onto the corresponding hyperplane of the target. For the rows $X$ the coordinates of the whole binary space are permuted and translated. On the value fibres we have plain translations. Hence in every sample in which $u$ holds, $b$ has exactly one preimage, and it is new. Integrating over the common sample gives the right-hand equality of~\eqref{eq:marg}. If the whole source fibre already lies in $U$, the bijective action on the fibre makes the event $u$ impossible for such a new $b$, and the sum is zero. Both arguments also hold for a parent map of probability zero.

\paragraph{Integer coefficients with the same support.}
The graph is finite. Hence for the function $R_p$ there is a power $\Delta_p$ of $p$ that clears all denominators at once. Since $\gcd(\Delta_2,\Delta_3)=1$, there are $\lambda_2,\lambda_3\in\Z$ with $\lambda_2\Delta_2+\lambda_3\Delta_3=1$. We put
\[
Q^t=\tfrac12\,(R_2+R_3),\qquad Z^t=\lambda_2\Delta_2R_2+\lambda_3\Delta_3R_3 .
\]
The equations~\eqref{eq:marg} are linear. Both functions take the value one at the empty map and at $t$. The function $Z^t$ is integral. If $Q^t$ vanishes at some map, so do the two nonnegative functions $R_p$, and hence also $Z^t$. Thus $\supp Z^t\subseteq\supp Q^t\subseteq M$. The map $t$ was arbitrary and $M$ is common to all $t$. Lemma~\ref{lem:cert} proves $A\eqZ B$ for the coloured graphs of the construction.

\section{Removing the colours without increasing the depth}\label{sec:colours}

We number the ten colours as $i=1,\dots,10$ and put $q_i=i+1$. To every original vertex $x$ of colour $i$ we attach $q_i$ private leaves, and we forget the colours. We write $A_k$, $B_k$ for the new graphs. The leaves have degree one, while every original vertex has degree at least two. The number of leaves adjacent to an original vertex recovers its colour exactly. Every isomorphism of the new graphs would restrict to a colour-preserving isomorphism of the original ones. Hence $A_k\not\cong B_k$.

It remains to transfer the witnesses to the same $k$. For the purposes of the proof we write the vertices over $x$ as $(x,0)$ for the original vertex and $(x,j)$ for the $j$-th leaf, $1\le j\le q_i$. The second index is an auxiliary numbering; the graph itself has only the adjacency relation. The family $\widetilde M$ consists of the maps $s$ of size at most $k$ that preserve the second index and for which the pairs $(x,y)$ with $s(x,j)=(y,j)$ define an injective map $\bar s$ of the bases with $\bar s\in M$. This family is closed under restriction. It preserves and reflects the edges between original vertices, because $\bar s$ does. Adjacency between an original vertex and a leaf is exactly equality of the bases, and there is no edge between two leaves. Hence every $s\in\widetilde M$ is a partial isomorphism.

For $t\in\widetilde M$ we take the two functions of $\bar t$. For each of them we define
\[
\widetilde w_s=\begin{cases} w_{\bar s}, & s\in\widetilde M,\\ 0, & \text{otherwise.}\end{cases}
\]
For a new vertex $(x,j)$ of the source there are two cases. If $x$ already lies in $\dom\bar s$, the only compatible image is $(\bar s(x),j)$. It exists because the bases have the same colour, and it is available because $\bar s$ is injective and $(x,j)$ has not yet been mapped. The new projection equals $\bar s$, so the marginal equation holds. If $x$ is new, the admissible terms of the extension correspond one to one to the admissible extensions $\bar s\cup\{x\mapsto y\}$ via the image $(y,j)$. The other terms have coefficient zero, and the equation is the original source equation. The same argument, starting from a new $(y,j)$ in the target, gives the target equation. If $s\notin\widetilde M$, no extension of $s$ lies in $\widetilde M$, by closure under restriction. This also covers all zero terms.

The empty map and $t$ still have coefficient one. Integrality and the inclusions of supports transfer pointwise. Lemma~\ref{lem:cert} gives $A_k\eqZ B_k$. Since $k\ge1$ was arbitrary and the signature now consists of the adjacency relation only, this completes the proof of Theorem~\ref{thm:main}. \qed

\section{How large must \texorpdfstring{$k$}{k} be?}\label{sec:width}

For even $n=2m\ge6$ let $A(n)$, $B(n)$ be the coloured graphs of Section~\ref{sec:graphs} with index set $D=[n]$, and let $A^\circ(n)$, $B^\circ(n)$ be their uncoloured versions from Section~\ref{sec:colours}. Thus $A_k=A^\circ(n)$ and $B_k=B^\circ(n)$ for the value of $n$ chosen in Section~\ref{sec:param}. The rows $X$ alone give $A^\circ(n)$ at least $\binom n2 2^{2(n-2)}$ vertices. Each of the ten colours has $2^{O(n)}$ vertices, each with a bounded number of leaves. So the number $N(n)$ of vertices of $A^\circ(n)$ satisfies $\log N(n)=\Theta(n)$.

\begin{theorem}\label{thm:width}
Let $n=2m\ge6$ be even and $K\ge\min\{\lceil\log_2n\rceil+2,\ \lceil\log_4(n-2)\rceil+3\}$. Then $A(n)\not\equiv^\Z_K B(n)$ and $A^\circ(n)\not\equiv^\Z_KB^\circ(n)$. Conversely, if (P1)--(P4) hold with $d$ replaced by some $k\ge1$, then $A(n)\equiv^\Z_{k'}B(n)$ and $A^\circ(n)\equiv^\Z_{k'}B^\circ(n)$ for all $k'\le k$. This is the case for $k=\lfloor\log_4n\rfloor-1$ whenever $n\ge24$.
\end{theorem}

Let $d_\Z(n)$ be the least $K$ with $A(n)\not\equiv^\Z_KB(n)$, and define $d^\circ_\Z(n)$ in the same way for the uncoloured graphs. By the theorem, both numbers lie between $\lfloor\log_4n\rfloor$ and $\log_4n+4$ for every even $n\ge24$, and both are $\Theta(\log\log N(n))$. For $n=80$, the case $k\le3$, $m=40$ of Section~\ref{sec:param} gives $4\le d_\Z(80),d^\circ_\Z(80)\le7$. In terms of size, suppose that $A^\circ(n)\equiv^\Z_kB^\circ(n)$. Then $k<\log_4n+4$, so $n>4^{k-4}$, and the graphs have $2^{\Omega(4^k)}$ vertices. Conversely, the least even $n\ge\max\{24,4^{k+1}\}$ gives an $\equiv^\Z_k$-equivalent pair with $2^{O(4^k)}$ vertices. Within this family, therefore, size doubly exponential in $k$ is necessary.

\paragraph{Idea of the upper bound.}
We pin a few rows whose images determine, for every surviving extension, the permutation $g$ of $D$ that it induces. The integral extensions of the pinned section then force $g$ to be both even and odd, exactly as an isomorphism would in Section~\ref{sec:graphs}.

\begin{proof}
\emph{Lower bounds.} The proof of Theorem~\ref{thm:main} in Sections~\ref{sec:cert}--\ref{sec:colours} uses only (P1)--(P4), applied with $k\le d$. If they hold with $d$ replaced by $k$, they also hold with $d$ replaced by any $k'\le k$, since their right-hand sides increase with $d$. So the proof applies to every $k'\le k$. Now let $n\ge24$ and $k=\lfloor\log_4n\rfloor-1$. Then $k\ge1$ and $m\ge2\cdot4^k$. (P3) is clear, and (P4) follows from $2\cdot4^k>4k+3$. (P1) holds because $4^k\le n/4$ and $3n/4\ge3k+2$. For $k\ge2$, (P2) follows from $2\cdot4^k>3(k+\binom k2)+3k+5$. For $k=1$, (P2) says $m>11$, which holds because $n\ge24$.

\emph{Reduction to the coloured graphs.} Suppose that the fixpoint $\mathcal S$ of Section~\ref{sec:cert} at width $K$ is nonempty, either for $A(n),B(n)$ or for $A^\circ(n),B^\circ(n)$. We use only three of its properties.
\begin{enumerate}[label=(\alph*)]
\item $\mathcal S$ is closed under restriction and consists of partial isomorphisms of size at most $K$.
\item Every $t\in\mathcal S$ with $|t|<K$ has the bijective forth property in $\mathcal S$.
\item Every $t\in\mathcal S$ is $\Z$-extendable in $\mathcal S$.
\end{enumerate}
Iterating (b) from the empty section, every set of at most $K$ source vertices is the domain of a section in $\mathcal S$. Consider the uncoloured case and let $\{v\mapsto w\}\in\mathcal S$; note that $K\ge3$. The bijection $f$ given by (b) maps the neighbours of $v$ onto those of $w$. By (a), $\{u\mapsto f(u)\}\in\mathcal S$ for every $u$, and the same argument applied to these sections shows that $f$ preserves degrees. Hence sections of size one preserve the degree and the number of neighbours of degree one. In $A^\circ(n)$ and $B^\circ(n)$ the vertices of degree one are exactly the leaves, every original vertex has degree at least two, and an original vertex of colour $i$ is adjacent to exactly $q_i$ leaves. Therefore every section in $\mathcal S$ maps leaves to leaves and original vertices to original vertices of the same colour. The sections of $\mathcal S$ whose domain consists of original vertices then satisfy (a)--(c) for the coloured graphs $A(n)$, $B(n)$. Indeed, the bijections in (b) map original vertices onto original vertices, and in (c) the integral sections over sets of original vertices are supported on such maps. So it suffices to derive a contradiction from a family $\mathcal S$ with (a)--(c) for $A=A(n)$ and $B=B(n)$.

\emph{The pin.} We fix a set $C_0$ of source vertices in one of two ways, using the one with $|C_0|+2\le K$.
\begin{enumerate}[label=(\roman*)]
\item Let $\rho=\lceil\log_2n\rceil$. The $2^\rho\ge n$ words in $\{0,1\}^\rho$ form $2^{\rho-1}\ge m$ pairs of complementary words. Fix a bijection $w$ from $D$ onto the union of $m$ of these pairs. For each position $\pi\in[\rho]$ the set $S_\pi=\{l\in D:w(l)_\pi=1\}$ then has exactly $m$ elements. For each $\pi$ choose $\xi^\pi\in\F_3^{D\setminus S_\pi}$ with $\sum_{j\notin S_\pi}\xi^\pi_j=\beta_{S_\pi}$, so that $Y_{S_\pi,\xi^\pi}$ is a vertex of $A$. Let $C_0=\{Y_{S_\pi,\xi^\pi}:\pi\in[\rho]\}$, so $|C_0|\le\rho$.
\item Let $\rho=\lceil\log_4(n-2)\rceil$ and fix $e_0=\{i,j\}$ with $i<j$. Fix an injective map $w$ from $D\setminus e_0$ to $(\F_2^2)^\rho$. For $\pi\in[\rho]$ let $x^\pi\in\F_2^{N(e_0)}$ be the vector with $(x^\pi_{\{i,l\}},x^\pi_{\{j,l\}})=w(l)_\pi$ for all $l\notin e_0$. Let $C_0$ consist of $X_{e_0,0}$ and the rows $X_{e_0,x^\pi}$, $\pi\in[\rho]$, so $|C_0|\le\rho+1$.
\end{enumerate}
Let $t_0\in\mathcal S$ be a section with domain $C_0$. All sections used below have at most $|C_0|+2\le K$ elements, and (b) is applied only to sections with at most $|C_0|+1<K$ elements.

\emph{Claim 1. There is a permutation $g$ of $D$ such that $s(l)=g(l)$ for every $s\in\mathcal S$ with $t_0\subseteq s$ and every $l\in D\cap\dom s$.}

In case (i) let $S'_\pi$ be the centre of the row $t_0(Y_{S_\pi,\xi^\pi})$. For $\tilde l\in D$ let $w'(\tilde l)\in\{0,1\}^\rho$ record, for each $\pi$, whether $\tilde l\in S'_\pi$. If $s\supseteq t_0$ maps $l$ to $\tilde l$, then $w'(\tilde l)=w(l)$, because $s$ preserves adjacency between $l$ and the pinned rows. By (b) there is a bijection $f$ extending $t_0$ with $t_0\cup\{v\mapsto f(v)\}\in\mathcal S$ for every $v$. It maps $D$ onto $D$, and $w'\circ f=w$ on $D$. Since $w$ is injective, so is $w'$, and $g=f|_D$ is the only possible image.

In case (ii) write $t_0(X_{e_0,0})=X_{e',y^0}$ and $t_0(X_{e_0,x^\pi})=X_{e'_\pi,y^\pi}$. By (b) some extension of $t_0$ in $\mathcal S$ maps $P_{e_0}$. Every vertex of $X$ has exactly one neighbour in $P$, namely the vertex of its centre. So this image is $P_{e'}$, and $e'_\pi=e'$ for all $\pi$. Since $c_{e_0}=0$, the row $X_{e_0,0}$ of $A$ is adjacent to $O_{ij}$. Write $e'=\{i',j'\}$ so that the neighbour of $X_{e',y^0}$ in $O$ is $O_{i'j'}$. Let $s\supseteq t_0$ map $l\in D$ to $\tilde l$. By (a), the section $t_0\cup\{l\mapsto\tilde l\}$ lies in $\mathcal S$, and by (b) it extends to $P_{e_0}$, with image $P_{e'}$, and also to $O_{ij}$, with image $O_{i'j'}$, the only neighbour of $X_{e',y^0}$ in $O$. Within $D$, the vertex $P_{e_0}$ is adjacent exactly to $i$ and $j$, and $O_{ij}$ exactly to $i$. Hence $\tilde l\in e'$ if and only if $l\in e_0$, and $\tilde l=i'$ if and only if $l=i$. So the images $i'$ of $i$ and $j'$ of $j$ are forced, and $\tilde l\notin e'$ for $l\notin e_0$.

Now let $l\notin e_0$, so that $\tilde l\notin e'$, and let $\eta\in\{i,j\}$ with forced image $\eta'$. By (b), $t_0\cup\{l\mapsto\tilde l\}$ extends to the vertex $P_{\{\eta,l\}}$, whose image $P_f$ then satisfies $\tilde l\in f$. Restrict this section to $t_0\cup\{P_{\{\eta,l\}}\mapsto P_f\}$ and extend it by (b) to $\eta$. The image of $\eta$ is $\eta'$, so $\eta'\in f$ and $f=\{\eta',\tilde l\}$. Extend the same restriction by (b) to the vertex $H=H_{e_0\{\eta,l\},0}$, which is adjacent to $X_{e_0,0}$ and to $P_{\{\eta,l\}}$. The image of $H$ is adjacent to $X_{e',y^0}$ and to $P_{\{\eta',\tilde l\}}$, so it is $H_{e'\{\eta',\tilde l\},b}$ with $b=y^0_{\{\eta',\tilde l\}}$. For every $\pi$, the vertex $H$ is adjacent to $X_{e_0,x^\pi}$ if and only if $x^\pi_{\{\eta,l\}}=0$. Its image is adjacent to $X_{e',y^\pi}$ if and only if $y^\pi_{\{\eta',\tilde l\}}=y^0_{\{\eta',\tilde l\}}$. Hence $x^\pi_{\{\eta,l\}}=y^\pi_{\{\eta',\tilde l\}}+y^0_{\{\eta',\tilde l\}}$. Define $w'(\tilde l)\in(\F_2^2)^\rho$ from $t_0$ alone by
\[
w'(\tilde l)_\pi=\bigl(y^\pi_{\{i',\tilde l\}}+y^0_{\{i',\tilde l\}},\ y^\pi_{\{j',\tilde l\}}+y^0_{\{j',\tilde l\}}\bigr).
\]
Taking $\eta=i$ and $\eta=j$ gives $w'(\tilde l)=w(l)$. As in case~(i), the bijection given by (b) at $t_0$ shows that $w'$ is injective on $D\setminus e'$, and the claim follows.

\emph{Claim 2. For every vertex $v$ of $A$ and every $s\in\mathcal S$ with $t_0\subseteq s$ and $v\in\dom s$, the vertex $s(v)$ lies in the target fibre that Table~\ref{tab:shifts} assigns to $v$ for the permutation $g$.}

It suffices to consider $s=t_0\cup\{v\mapsto s(v)\}$. By (b) we may add any vertex $u$ of $D$. Its image is $g(u)$ by Claim~1, and $u$ is adjacent to $v$ exactly when $g(u)$ is adjacent to $s(v)$. Hence the neighbourhood of $s(v)$ in $D$ is the image under $g$ of that of $v$. This settles $D^*$, $P$, $L^\nu$ and $Y$; for instance, $l^*$ is the only vertex of $D^*$ adjacent to $l$, so its image is $g(l)^*$. Adding $b^*$ instead, we see that the image of $O_{ab}$ is adjacent to $g(a)$ and to $g(b)^*$, so it is $O_{g(a)g(b)}$. Adding $P_e$ and $P_f$, the image of a vertex $H_{ef,b}$ lies over $\{ge,gf\}$. Adding $P_e$, the image of $X_{e,x}$ is a row over $ge$.

\emph{Integral measures.} By (c) there are elements $r_W\in\Z\mathcal S(W)$, for all sets $W$ of at most $K$ source vertices, that are compatible under restriction and satisfy $r_{C_0}=[t_0]$. For $W\supseteq C_0$ let $r^0_W$ be the sum of those terms $r_W(s)[s]$ of $r_W$ with $s|_{C_0}=t_0$. For $C_0\subseteq W'\subseteq W$ we have $s|_{C_0}=t_0$ if and only if $(s|_{W'})|_{C_0}=t_0$. So restriction maps $r^0_W$ to $r^0_{W'}$, and the coefficients of $r^0_W$ add up to those of $r^0_{C_0}=[t_0]$, that is, to one. For a set $U$ of at most two source vertices and a map $\varphi$ from $U$ to $B$ put
\[
\mu_U(\varphi)=\sum\bigl\{r^0_{C_0\cup U}(s):s|_U=\varphi\bigr\}\in\Z .
\]
Then $\sum_\varphi\mu_U(\varphi)=1$, and $\mu_{U'}$ is the marginal of $\mu_U$ for $U'\subseteq U$. Every $\varphi$ with $\mu_U(\varphi)\ne0$ is the restriction of a section $s\in\mathcal S$ with $t_0\subseteq s$, so Claims~1 and~2 apply to it. For a prime $p$ and a function $F$ from such maps to $\F_p$ we write $E[F]=\sum_\varphi\mu_U(\varphi)F(\varphi)\in\F_p$. If $F$ depends only on $\varphi|_{U'}$, the value computed over $U$ equals the one computed over $U'$. If $F$ takes the same value $c$ wherever $\mu_U\ne0$, then $E[F]=c$.

\emph{The part over $\F_3$.} For a source vertex $L^\nu_{T,a}$ let $M^\nu_T(a)=E[\val]$, where $\val(\varphi)\in\F_3$ is the value of the image of $L^\nu_{T,a}$; by Claim~2 this image is a vertex over $gT$. The images of the adjacent pairs $L^0_{T,a}L^1_{T,a}$, $L^1_{T,a}L^2_{T,a}$ and $L^0_{T,a}L^2_{T,a+1}$ are adjacent, so their values are equal, equal, and differ by one, respectively. Hence $M^0_T(a)=M^1_T(a)=M^2_T(a)$ and $M^2_T(a+1)=M^0_T(a)+1$. Therefore $M^0_T(a)=a+h_{gT}$ with $h_{gT}=M^0_T(0)$, which defines $h\in V_{m+1}$. For $|S|=m$ choose a row $Y_{S,x}$ of $A$. Its images are rows $Y_{gS,y}$ of $B$, so $\sum_{j\notin S}y_{gj}=\beta_{gS}$. For $j\notin S$ the row is adjacent to $L^0_{S+j,x_j}$. Moreover, $Y_{gS,y}$ is adjacent to exactly one vertex of $L^0$ over $g(S+j)$, the one with value $y_{gj}$. Therefore $E[y_{gj}]=M^0_{S+j}(x_j)=x_j+h_{g(S+j)}$, and summing over $j\notin S$ gives $\beta_{gS}=\beta_S+(\partial h)_{gS}$. Thus $\partial h=\beta-g\beta$, and $g$ is even by Section~\ref{sec:graphs}.

\emph{The part over $\F_2$.} For an edge $ef$ of $J$ let $a_{ef}=a_{fe}=E[\val]$ for the source vertex $H_{ef,0}$, whose images lie over $\{ge,gf\}$. Let $e=\{i_1,i_2\}$ with $i_1<i_2$. The row $X_{e,0}$ of $A$ is adjacent to $H_{ef,0}$ for all $f\sim e$, and to $O_{i_1i_2}$ because $c_e=0$. Its images are rows $X_{ge,y}$. Adjacency to the image of $H_{ef,0}$ gives $E[y_{gf}]=a_{ef}$. The image of $O_{i_1i_2}$ is $O_{g(i_1)g(i_2)}$. By the definition of $B$, adjacency to it means that $\sum_{f\sim e}y_{gf}=c'_{ge}+\epsilon_e(g)$. Hence $\sum_{f\sim e}a_{ef}=c'_{ge}+\epsilon_e(g)$ for every vertex $e$ of $J$. Summing over all $e$, every $a_{ef}$ occurs twice, and we obtain $0=1+\inv(g)$ modulo $2$. So $g$ is odd, a contradiction.
\end{proof}

\begin{remark}
Lemma~\ref{lem:sign} needs two unprotected indices with the same profile, and $k$ observed vertices produce at most $4^k$ profiles. Case~(ii) of the proof shows that this count is attained. Once $4^\rho\ge n-2$, there are $\rho$ rows over a common centre that give the $n-2$ indices outside the centre pairwise different profiles. The pair is then distinguished at width $\rho+3$.
\end{remark}

\section{A polynomial-time invariant}\label{sec:invariant}

To see that the pairs of Theorem~\ref{thm:main} are easy instances for isomorphism testing, we vary the charges. For $b\in V_m$ and $c\in\F_2^{V(J)}$ let $G(b,c)$ be the coloured graph defined as in Section~\ref{sec:graphs}, with two changes. First, the rows over $S$ are the vertices $Y_{S,x}$ with $\sum_{j\notin S}x_j=b_S$. Second, for $e=\{i,j\}$ with $i<j$, the row $X_{e,x}$ is adjacent to $O_{ij}$ if $\sum_fx_f=c_e$ and to $O_{ji}$ otherwise. Let $G^\circ(b,c)$ be its uncoloured version from Section~\ref{sec:colours}. We call these graphs \emph{templates}; thus $A=G(\beta,0)$ and $B=G(\beta,c')$. Put $p(c)=\sum_ec_e\in\F_2$ and $Q=V_m/\partial V_{m+1}$. Recall from Section~\ref{sec:graphs} that $[\beta]\ne0$ and $g[\beta]=\sgn(g)[\beta]$ in $Q$.

\begin{lemma}[Lift test]\label{lem:lift}
Let $b,\hat b\in V_m$ and $c,\hat c\in\F_2^{V(J)}$. The following are equivalent: $G(b,c)\cong G(\hat b,\hat c)$; $G^\circ(b,c)\cong G^\circ(\hat b,\hat c)$; there is a permutation $g$ of $D$ with $[\hat b]=[gb]$ in $Q$ and $\sgn(g)=(-1)^{p(c)+p(\hat c)}$.
\end{lemma}

\begin{proof}
The leaves recover the colours, as in Section~\ref{sec:colours}, which gives the first equivalence. For the second, the computation of Section~\ref{sec:graphs} applies verbatim. An isomorphism induces a permutation $g$ of $D$ and shifts $h\in V_{m+1}$ and $a\in\F_2^{E(J)}$ with
\[
\partial h=\hat b-gb,\qquad \sum_{f\sim e}a_{ef}=c_e+\hat c_{ge}+\epsilon_e(g)\quad(e\in V(J)).
\]
Conversely, every solution of these equations defines an isomorphism. The first equation is solvable exactly when $[\hat b]=[gb]$. Since $J$ is connected, the second is solvable exactly when its right-hand sides have sum zero, that is, when $p(c)+p(\hat c)+\inv(g)=0$.
\end{proof}

\begin{proposition}\label{prop:inv}
Let $\mathcal T$ be the class of graphs isomorphic to a template $G^\circ(b,c)$ with $[b]\in\F_3[\beta]$. For such a graph write $[b]=\lambda[\beta]$ and put
\[
I=(-1)^{p(c)}\lambda\in\F_3 .
\]
\begin{enumerate}[label=(\roman*)]
\item The value $I$ does not depend on the choice of the template. Two graphs in $\mathcal T$ with the same number of vertices are isomorphic if and only if they have the same value of $I$.
\item $I(A_k)=1$ and $I(B_k)=-1$ for every $k$.
\item Given a graph in $\mathcal T$, its value $I$ can be computed in time polynomial in its number of vertices.
\end{enumerate}
\end{proposition}

\begin{proof}
Since $[\beta]\ne0$, the coefficient $\lambda$ is determined by $b$. Every centre $S$ carries $3^{m-1}$ rows and every centre $e$ carries $4^{n-2}$ rows, whatever $b$ and $c$ are. Counting the leaves of Section~\ref{sec:colours}, every template $G^\circ(b,c)$ with $|D|=n=2m$ therefore has exactly
\[
N(n)=7n+5\tbinom n2+6n(n-1)+14(n-2)\tbinom n2+8\cdot4^{n-2}\tbinom n2+90\tbinom n{m+1}+12\cdot3^{m-1}\tbinom nm
\]
vertices. As $n=2m$ ranges over the positive even integers, every displayed term is nondecreasing in $n$, and the fifth one is strictly increasing. Hence two templates with the same number of vertices have the same $n$.

(i) Suppose that $G^\circ(b,c)\cong G^\circ(\hat b,\hat c)$ and $[b]=\lambda[\beta]$. Lemma~\ref{lem:lift} gives $g$ with $[\hat b]=[gb]=\lambda\sgn(g)[\beta]$ and $\sgn(g)=(-1)^{p(c)+p(\hat c)}$. Hence $\hat\lambda=\lambda\sgn(g)$ and $(-1)^{p(\hat c)}\hat\lambda=(-1)^{p(c)}\lambda$. Conversely, suppose that the two values agree. Let $g$ be the identity if $p(c)=p(\hat c)$, and a transposition otherwise. Then $[gb]=\sgn(g)\lambda[\beta]=(-1)^{p(c)+p(\hat c)}\lambda[\beta]=\hat\lambda[\beta]=[\hat b]$, and Lemma~\ref{lem:lift} gives an isomorphism.

(ii) We have $A_k=G^\circ(\beta,0)$ and $B_k=G^\circ(\beta,c')$ with $p(c')=1$.

(iii) Let $G$ be the input, with $N$ vertices. The computation has three steps.

\emph{Step 1: roles and a chart.} The vertices of degree one are the leaves, and the number of leaves at any other vertex gives its colour. Number $D$ arbitrarily. In every fibre of $L^0$ choose a vertex of value $0$, and number the fibre by the successor map, which goes from $L^0$ to $L^2$, to $L^1$ and back to $L^0$ around the nine-cycle; the values in $L^1$ and $L^2$ follow along the matchings. In every fibre of $H$ choose a vertex of value $0$. All other indices are read off from the neighbours in $D$, $D^*$ and $P$.

\emph{Step 2: the charges.} Read the coordinates of the rows from their neighbours in $L^0$ and $H$. Suppose that $G\cong G^\circ(b,c)$. These coordinates then differ from those of $G^\circ(b,c)$ by a permutation of $D$ and by translations in the value fibres. By the computation in the proof of Lemma~\ref{lem:lift}, in these coordinates $G$ is a template $G^\circ(\tilde b,\tilde c)$. Here $\tilde b_S$ is the common sum of the coordinates of the rows over $S$. For $e=\{i,j\}$ with $i<j$ in the chosen numbering, $\tilde c_e$ is the value of $\sum_fx_f$ for which the rows over $e$ are attached to $O_{ij}$.

\emph{Step 3: the coefficient.} By~(i), $I(G)=(-1)^{p(\tilde c)}\tilde\lambda$, where $\tilde\lambda$ is the unique element of $\F_3$ with $\tilde b-\tilde\lambda\beta\in\partial V_{m+1}$. We find it by testing the three candidates with Gaussian elimination over $\F_3$, in a matrix with $\binom nm$ rows and $\binom n{m+1}$ columns; both numbers are at most $N$.

Each step takes time polynomial in $N$.
\end{proof}

\begin{remark}
It is routine to decide in polynomial time whether a given graph is isomorphic to a template, and to compute its $n$: one recovers the roles and the coordinates as in~(iii) and checks the incidence pattern of Section~\ref{sec:graphs} fibre by fibre. With Proposition~\ref{prop:inv}, this shows that the class of graphs isomorphic to some $A_k$ is decidable in polynomial time. A sentence defining this class holds in every $A_k$ and in no $B_k$. Consequently, suppose that every sentence of a logic has the same truth value on $A_k$ and on $B_k$ for some $k$. Then that logic does not capture polynomial time on graphs. By Theorem~\ref{thm:main}, for every $k$ the relation $\eqZ$ does not separate the two values of $I$ on $A_k$ and $B_k$.
\end{remark}

\section{The graphs of Berkholz and Grohe at width two}\label{sec:bg}

We recall the construction of~\cite[Sections~3 and~4]{BG17}. Let $\Gamma$ be a finite abelian group. A coset constraint $C=((x_1,\dots,x_\kappa),\Delta_C+\gamma_C)$ consists of a tuple of variables and a coset of a subgroup $\Delta_C\le\Gamma^\kappa$. For a system $\mathcal C$ of such constraints, the graph $G(\mathcal C)$ has the following vertices and edges.
\begin{itemize}
\item For every variable $x$ and every $\gamma\in\Gamma$, a vertex $\gamma^{(x)}$ of colour $L^{(x)}$.
\item For every constraint $C$ and every $\alpha\in\Delta_C+\gamma_C$, a vertex $\alpha^{(C)}$ of colour $L^{(C)}$.
\item For every such $\alpha$ and every $i\in[\kappa]$, an edge of colour $M(i)$ between $\alpha^{(C)}$ and $\alpha_i^{(x_i)}$.
\end{itemize}
The homogeneous graph $\widetilde G(\mathcal C)$ is defined in the same way, with every coset $\Delta_C+\gamma_C$ replaced by $\Delta_C$ and with the same colours. For graphs $G_1$, $G_2$, the sequence graph $\langle G_1,G_2\rangle$ is their disjoint union with two further vertices $v_1$, $v_2$. The vertex $v_1$ is adjacent to all vertices of $G_1$, and $v_2$ to all vertices of $G_1$ and $G_2$. For pairs $(G_1^0,G_1^1)$ and $(G_2^0,G_2^1)$, the or-construction is the pair $(G^0,G^1)$ in which $G^\epsilon$ is the disjoint union of the graphs $\langle G_1^{a_1},G_2^{a_2}\rangle$ with $a_1+a_2\equiv\epsilon\pmod2$.

In the proof of~\cite[Theorem~4.1]{BG17}, $\Gamma=\Z_2\times\Z_3$, $\iota_2=(1,0)$ and $\iota_3=(0,1)$, and for every level $\ell$ an oriented $3$-regular expander is chosen according to $\ell$. For $p\in\{2,3\}$ the system $\mathcal C_{\ell,p}$ consists of two parts. The first is a Tseitin system over this expander, whose right-hand side at one fixed vertex is replaced by a new variable $x^*$. The second is the unary constraint $(x^*,\{\iota_p\})$. The pair of level $\ell$ is
\[
(G_\ell,\widetilde G_\ell)=\bigl(G(\mathcal C_{\ell,2}),\widetilde G(\mathcal C_{\ell,2})\bigr)\vee\bigl(G(\mathcal C_{\ell,3}),\widetilde G(\mathcal C_{\ell,3})\bigr).
\]
So $G_\ell$ consists of $\langle G(\mathcal C_{\ell,2}),G(\mathcal C_{\ell,3})\rangle$ and $\langle\widetilde G(\mathcal C_{\ell,2}),\widetilde G(\mathcal C_{\ell,3})\rangle$, and $\widetilde G_\ell$ of the two mixed sequence graphs. The level-$\ell$ Diophantine relaxation of isomorphism has an integral solution for $(G_\ell,\widetilde G_\ell)$~\cite[Theorem~4.1]{BG17}. We use only two properties of $\mathcal C_{\ell,2}$ and $\mathcal C_{\ell,3}$. First, they have the same variables and the same subgroups $\Delta_C$. Second, both are unsatisfiable~\cite[proof of Theorem~4.1]{BG17}: adding the constraints at all vertices of the expander, every edge variable cancels, which forces $x^*=0$.

\begin{proposition}\label{prop:bg}
For every level $\ell$, the vertex- and edge-coloured graphs $G_\ell$, $\widetilde G_\ell$ satisfy $G_\ell\not\equiv^\Z_k\widetilde G_\ell$ for every $k\ge2$, and $G_\ell\equiv^\Z_1\widetilde G_\ell$.
\end{proposition}

\begin{proof}
Fix $\ell$, and write $G$, $\widetilde G$, $\mathcal C_2$, $\mathcal C_3$ for $G_\ell$, $\widetilde G_\ell$, $\mathcal C_{\ell,2}$, $\mathcal C_{\ell,3}$. Let $k\ge2$, and suppose that the fixpoint $\mathcal S$ at width $k$ is nonempty. We use properties (a)--(c) from the proof of Theorem~\ref{thm:width}. Adjacency means adjacency by an edge of any colour.

\emph{Roles.} As in the proof of Theorem~\ref{thm:width}, every section of size one in $\mathcal S$ preserves the degree and the number of neighbours of each degree. Let $N_0$ be the number of vertices of $G(\mathcal C_2)$. By the first property, $N_0$ is also the number of vertices of $G(\mathcal C_3)$, $\widetilde G(\mathcal C_2)$ and $\widetilde G(\mathcal C_3)$. In every sequence graph, $v_1$ has degree $N_0$ and $v_2$ has degree $2N_0$, while every other vertex has degree less than $N_0$. Indeed, inside its copy of $G(\mathcal C_p)$ or $\widetilde G(\mathcal C_p)$, a vertex $\gamma^{(x)}$ has at most $\sum_C|\Delta_C|\le N_0-|\Gamma|$ neighbours, and a vertex $\alpha^{(C)}$ has at most as many as $C$ has variables; each of them has at most two further neighbours, $v_1$ and $v_2$. Hence sections of size one in $\mathcal S$ map vertices $v_1$ to vertices $v_1$, and vertices $v_2$ to vertices $v_2$. A vertex of a copy of $G(\mathcal C_p)$ or $\widetilde G(\mathcal C_p)$ lies in the first position of its sequence graph if and only if it has a neighbour of degree $N_0$, so these sections also preserve positions. Being partial isomorphisms, they preserve the colours $L^{(x)}$ and $L^{(C)}$.

\emph{One component carries mass one.} Let $a$ be the vertex $v_2$ of $\langle\widetilde G(\mathcal C_2),\widetilde G(\mathcal C_3)\rangle$. By (b) at the empty section there is a section $\{a\mapsto b\}\in\mathcal S$, and $b$ is the vertex $v_2$ of a sequence graph $\mathcal J$ of $\widetilde G$, which is mixed. So there is a position $j_0$ in which $\mathcal J$ has $G(\mathcal C)$, where $\mathcal C=\mathcal C_2$ if $j_0=1$ and $\mathcal C=\mathcal C_3$ if $j_0=2$, while the source has $\widetilde G(\mathcal C)$ in this position. By (c) there are compatible elements $r_W\in\Z\mathcal S(W)$, $|W|\le k$, with $r_{\{a\}}=[a\mapsto b]$. For a source vertex $z$ and a target vertex $z'$ let $\mu_z(z')$ be the coefficient of $\{z\mapsto z'\}$ in $r_{\{z\}}$. Let $z$ be adjacent to $a$. Every section in the support of $r_{\{z,a\}}$ maps $z$ and $a$ to adjacent vertices, which lie in the same connected component. Restricting $r_{\{z,a\}}$ to $z$ and to $a$ therefore gives
\[
\sum_{z'\in\mathcal J}\mu_z(z')=\sum_{a'\in\mathcal J}\mu_a(a')=1 .
\]
Since $a$ is adjacent to all vertices of the two graphs in its sequence graph, this holds for all of them.

\emph{A solution of $\mathcal C$.} For a variable $x$ of $\mathcal C$ let $z_x$ be the vertex $0^{(x)}$ of the source copy of $\widetilde G(\mathcal C)$ in position $j_0$. For a constraint $C$ let $z_C$ be the vertex $0^{(C)}$ there, which exists because $0\in\Delta_C$. By the first step, every $z'\in\mathcal J$ with $\mu_{z_x}(z')\ne0$ is a vertex $\gamma^{(x)}$ of the copy of $G(\mathcal C)$ in $\mathcal J$. Likewise, every $z'\in\mathcal J$ with $\mu_{z_C}(z')\ne0$ is a vertex $\alpha^{(C)}$ of that copy, with $\alpha\in\Delta_C+\gamma_C$. Using these target vertices, put
\[
A_x=\sum_{\gamma\in\Gamma}\mu_{z_x}\bigl(\gamma^{(x)}\bigr)\,\gamma\in\Gamma,\qquad
A_C=\sum_{\alpha\in\Delta_C+\gamma_C}\mu_{z_C}\bigl(\alpha^{(C)}\bigr)\,\alpha\in\Gamma^\kappa .
\]
In each sum the coefficients add up to one. A coset is closed under integer combinations whose coefficients add up to one, so $A_C\in\Delta_C+\gamma_C$. Let $C$ have scope $(x_1,\dots,x_\kappa)$ and let $i\in[\kappa]$. The vertices $z_C$ and $z_{x_i}$ are joined by an edge of colour $M(i)$. Hence every section in the support of $r_{\{z_C,z_{x_i}\}}$ maps them to vertices $\alpha^{(C)}$ and $\alpha_i^{(x_i)}$ for some $\alpha$, and one of these lies in $\mathcal J$ if and only if the other does. We restrict $r_{\{z_C,z_{x_i}\}}$ to each of its two points, keeping only the sections that map into $\mathcal J$; this gives $(A_C)_i=A_{x_i}$. Thus $(A_x)_x$ satisfies every constraint of $\mathcal C$, which contradicts its unsatisfiability.

\emph{Width one.} The colour classes of $G$ and $\widetilde G$ have equal sizes, because $|\Delta_C+\gamma_C|=|\Delta_C|$. So every partial isomorphism $t$ of size at most one extends to a colour-preserving bijection $f$. The sections $[\emptyset]$ and $[u\mapsto f(u)]$ witness that $t$ is $\Z$-extendable in $I_1(G,\widetilde G)$, and $f^{-1}$ does the same for $t^{-1}$. For $t=\emptyset$, $f$ itself gives the bijective forth property. Hence $I_1(G,\widetilde G)$ is its own fixpoint.
\end{proof}

\begin{remark}
\begin{enumerate}[label=(\arabic*)]
\item \'O~Conghaile observes that the test of~\cite{BG17} appears similar to checking the $\Z$-extendability of the empty section~\cite[Section~4.4]{OC22}. For every $\ell$, the level-$\ell$ Diophantine relaxation has an integral solution for $(G_\ell,\widetilde G_\ell)$, but the $\Z$-extendability of a single one-point section fails already at width two. The image of the one vertex $a$ decides which of the two disjuncts the target has to satisfy. This is exactly what the construction of Theorem~\ref{thm:main} avoids (Section~\ref{sec:disc}).
\item Besides the edge colours $M(i)$, which only identify the position of a variable in a constraint, the proof uses only that sections of size one preserve the vertex colours and the roles of $v_1$ and $v_2$. Suppose that the variables in each scope are distinct, as for Tseitin constraints over a graph without loops. Then the edge colours can be omitted, and the vertex colours, together with fresh colours for $v_1$ and $v_2$, can be encoded by pendant leaves as in Section~\ref{sec:colours}. By the degree argument in the proof of Theorem~\ref{thm:width}, the proof then applies to the resulting uncoloured graphs. Theorem~4.1 of~\cite{BG17} is stated for $3$-regular graphs. We have not analysed the passage from the coloured graphs of its proof to $3$-regular graphs.
\end{enumerate}
\end{remark}

\section{Discussion}\label{sec:disc}

\paragraph{Why a non-local coupling.}
Definition~6 asks for integral extensions of every surviving local isomorphism, so in our approach an integral combination of the two branches must remain available after conditioning on any local isomorphism. If the images of a bounded set of vertices decided which branch applies, conditioning on them would leave a single branch, and the B\'ezout combination of Section~\ref{sec:dist} would no longer be available. For constraint satisfaction, Lichter and Pago show that cohomological $k$-consistency solves the constraint satisfaction problem of their Theorem~1, an or-construction of systems of linear equations over $\Z_2$ and $\Z_3$, for every $k\ge4$~\cite[Theorem~3]{LP25}. Proposition~\ref{prop:bg} shows the corresponding fact for the isomorphism pairs of Berkholz and Grohe: there the image of one vertex decides the branch, and width two suffices. In the present construction the branch is the parity of the permutation of $D$, and Lemma~\ref{lem:sign} shows that every local isomorphism of at most $k$ vertices is compatible with both parities.

\paragraph{Size.}
The binary part has $\binom n2\,2^{2(n-2)}$ row vertices and the ternary part has $\binom{n}{m}3^{m-1}$ row vertices; with $n=2m$ and $m\ge 4^{d}$ the graphs have $N=2^{\Theta(m)}=2^{2^{\Theta(k)}}$ vertices. Measured by the number $N$ of vertices, the pairs of Theorem~\ref{thm:main} have $k=\Theta(\log\log N)$, so the theorem says nothing about values of $k$ that grow faster than $\log\log N$. Theorem~\ref{thm:width} shows that this is a limit of the construction and not of the analysis. The growth has two sources. Lemma~\ref{lem:sign} needs more than $4^k$ unprotected indices, and every row reads a whole star, so a few rows over one centre already separate all indices. By contrast, the graphs of~\cite{BG17} for the Diophantine relaxation have size linear in the level.

\paragraph{Algorithmic status.}
By Proposition~\ref{prop:inv} the pairs are distinguished in polynomial time. The invariant $I$ couples the two parts through the parity of an arbitrary numbering of $D$, and by Theorem~\ref{thm:main} no fixed level of $\equiv^\Z$ detects it on the pairs $A_k$, $B_k$. The pairs are therefore of interest for logics and for consistency methods, not as hard instances for isomorphism algorithms.

\paragraph{Open questions.}
\begin{enumerate}[label=(\arabic*)]
\item Are there pairs of non-isomorphic graphs of size polynomial in $k$, or even linear in $k$, that are $\eqZ$-equivalent? By Theorem~\ref{thm:width}, this needs a different construction.
\item Does $\eqZ$ distinguish, for some fixed $k$, the $3$-regular graphs of~\cite[Theorem~4.1]{BG17}? For the coloured graphs of its proof, Proposition~\ref{prop:bg} gives $k=2$.
\item \'O~Conghaile notes that his algorithms are likely expressible in rank logic extended by a quantifier for the solvability of linear equations over $\Z$. He asks whether such a logic captures polynomial time (Question~16 in the arXiv version of~\cite{OC22}). Rank logic itself does not capture polynomial time~\cite{Lic23}. Does such an extension, or choiceless polynomial time, distinguish $A_k$ from $B_k$ for all $k$? By the Remark after Proposition~\ref{prop:inv}, a negative answer for a logic would show that it does not capture polynomial time.
\end{enumerate}

\section*{AI methodology statement}
The author conceived this work, chose its strategy and directed how AI tools were applied at each step. AI tools assisted with elaborating and checking detailed arguments, with the literature search, with the translation of the manuscript from Greek into English and its editing, and with writing and running the finite checks of Appendix~\ref{app:checks}. Independent AI-assisted reviews served as critical readers. The author assessed every claim and outcome before adopting it, and takes full responsibility for the content, including references and any errors.

\appendix
\section{Finite computational checks}\label{app:checks}

The following checks were run on small or explicit instances. They are sanity checks only and are not used in the proofs. The scripts, the logs of the reference runs and a README with the commands and the software versions are provided as ancillary files with this paper or are available from the author; a single command runs all checks.

\begin{enumerate}[label=A.\arabic*,leftmargin=*]
\item \emph{The class $[\beta]$.} For $m=1,\dots,6$ ($n\le12$), exact linear algebra over $\F_3$ gives $\partial^{\mathsf T}\beta=0$, $\langle\beta,\beta\rangle=2^m$ and $[\beta]\neq0$, and $g[\beta]=\sgn(g)[\beta]$ for $g=(1\,2)$, $(2\,3)$ (when $n\ge3$) and $(1\,2\,\cdots\,n)$. For $n\le8$, among all $n!$ permutations, equation~\eqref{eq:F3} is solvable exactly for the even ones. On random instances the identities for $\partial^2U$ and $\partial^3U$ used in Lemma~\ref{lem:F3} also hold.
\item \emph{Small graphs.} For $m=2,3$ ($n=4,6$; $200$ and $4332$ vertices) we built the coloured graphs of Section~\ref{sec:graphs} and used the isomorphism and automorphism routines of \textsf{bliss} through \textsf{python-igraph}. As expected, $A\not\cong B$. If two of the values $c_e$ are flipped instead of one, the graph is isomorphic to $A$; if three are flipped, it is not. The $\F_2$ parts of $A$ and $B$ are isomorphic. The permutations of $D$ induced by the automorphisms of $A$, of its $\F_3$ part and of its $\F_2$ part form exactly the alternating group. The uncoloured graphs of Section~\ref{sec:colours} behave in the same way: $A_k\not\cong B_k$, and flipping two values gives a graph isomorphic to $A_k$. For $m=3$ \textsf{bliss} did not finish on the uncoloured graphs directly; there we first deleted the vertices of degree one and coloured every remaining vertex by the number of deleted neighbours, a reduction that does not change the answer. These values of $n$ are below the bounds of Section~\ref{sec:param}, which are not used in Section~\ref{sec:graphs}.
\item \emph{Lemma~\ref{lem:F3}.} The construction of the proof, with its exclusions, was executed for $r=1,2,3$ at the smallest admissible values $m=12,21,33$, and at $m=40$ for $r=3$, with random and with adversarial prescriptions (nearby centres, shared columns, extra columns inside or next to stars): $280$ instances. Every output lies in $\ker\partial$ and takes all prescribed values, and each choice of $\alpha,\alpha'$ had at least $7$ available elements.
\item \emph{Lemma~\ref{lem:F2}.} The construction of the proof was executed at $n=2s+3q+4$ for $(s,q)\in\{(1,0),(1,1),(0,3),(3,0),(2,2),(3,3)\}$ and at $n=80$ with $s=q=3$: $2440$ instances, all of which gave solutions of all vertex equations that keep the prescriptions.
\item \emph{Lemma~\ref{lem:sign}.} The construction of the proof was executed on random local realisations of at most $k$ vertices for $(k,n)=(1,10)$, $(2,24)$, $(3,76)$ and $(3,80)$, including instances made only of rows: $1700$ instances. The images of $t$ never changed, all observed equations held for $g\tau$, and the parity changed.
\item \emph{Local systems of Section~\ref{sec:dist}.} For the same parameters, $920$ local systems along random chains $U_1\subset U_2\subset\cdots$, biased towards vertices that interact with $t$, were all consistent, and every restriction map between consecutive solution spaces was onto.
\item \emph{The functions $R_p$.} An independent implementation computed the functions $R_p$ of Section~\ref{sec:dist} exactly, with rational arithmetic, for reduced parameters ($n\le16$, $k\le3$, below the bounds of Section~\ref{sec:param}), including the transfer of Section~\ref{sec:colours}. All tested sums in~\eqref{eq:marg} were exact, $R_p(\emptyset)=R_p(t)=1$ in all cases, and every map of positive probability was a partial isomorphism.
\item \emph{The invariant of Section~\ref{sec:invariant}.} We built the templates $G(b,c)$ for $m=2,3$ ($n=4,6$), attached the leaves of Section~\ref{sec:colours} and renamed all vertices at random. We then computed $I$ from the resulting graph as in the proof of Proposition~\ref{prop:inv}(iii), in a random chart: a random numbering of $D$ and random origins in all value fibres. In five runs for each of $A$ and $B$ and each $m$, we obtained $I(A)=1$ and $I(B)=-1$. On $20$ random templates with $[b]\in\F_3[\beta]$, the value computed from the graph agreed with the value $(-1)^{p(c)}\lambda$ computed from $(b,c)$. A charge $b$ with $[b]\notin\F_3[\beta]$ was recognised as such. For $m=2$, nauty compared $24$ random coloured templates with $\lambda=0,1,2$, that is, $276$ pairs: two templates were isomorphic exactly when their values of $I$ agreed. For $m=3$ this comparison was not run.
\end{enumerate}
An independent reimplementation, using nauty instead of \textsf{bliss}, reproduced the isomorphism and automorphism checks of A.2 for $m=2$, the non-isomorphism and the parity behaviour of the two parts for $m=3$, and the checks of A.5 and A.6 at $n=80$ for $k\le3$.

Lemma~\ref{lem:cert} and the relation $\eqZ$ itself were not checked by machine, and the functions $R_p$ were not computed at the parameters of Section~\ref{sec:param}: the graphs of Theorem~\ref{thm:main} are far too large for a direct computation. Theorem~\ref{thm:width} and Proposition~\ref{prop:bg} were not checked by machine either.

\end{document}